
\magnification=\magstep1
\parindent=0pt
\hfuzz15pt
\def\sqr#1#2{{\vcenter{\vbox{\hrule height.#2pt
\hbox{\vrule width.#2pt height#1pt \kern#1pt\vrule width.#2pt}
\hrule height.#2pt}}}}
\def\square{\mathchoice\sqr34\sqr34\sqr{2.1}3\sqr{1.5}3}
\def\harp#1{\buildrel \rightharpoonup \over #1}
\centerline{\bf Some Classical Solutions of Relativistic Membrane Equations}
\centerline{\bf in 4 Space-Time Dimensions}
\vskip1truecm
\centerline{J Hoppe*}
\centerline{Isaac Newton Institute for Mathematical Sciences}
\centerline{20 Clarkson Road}
\centerline{Cambridge CB3 0EH}
\centerline{UK}
\vskip3truecm
\centerline{\bf Abstract}
\vskip1truecm
\centerline{Various reductions, and some solutions of the classical equations}
\centerline{of motion of a relativistic membrane are given.}
\vskip1truecm
\centerline{January 94}
\vskip10truecm
* Heisenberg Fellow

\ \ On leave of absence from the Institute for Theoretical
Physics, Karlsruhe University.
\vfill\eject
As there now exists quite a variety of non-trivial reformulations,
and simplifications, of the standard parametric `minimal hypersurface'
equations, different kinds of reductions, and insight, can be used to find
explicit solutions.  Static solutions of the hydrodynamic equations derived
in [1], e.g., correspond to
infinitely extended membranes of fixed shape, moving with
the velocity of light.  Alternatively, solutions may be derived by
viewing the non parametric $z$-equation (cp [2]) as a consistency condition
of a higher dimensional free (!) wave equation and a non-linear constraint.
Using the orthonormal light cone gauge [3], on the other hand, for a
non-compact membrane with one rotational symmetry, an algebraic
`shape-changing' solution is found.  Also, while the closed membrane
light
cone solution of [3] contracts to a point,  one
can write down solutions of the same equation without the above
point-singularity, by relaxing the $S^1$-symmetry.  Two classes of
symmetry reductions of the membrane equations to ordinary differential
equations are then deduced, as well as the classical equivalence of
axially-symmetric membranes with strings in a given curved 3-dimensional
background.  At the end an idea is sketched which indicates a possible
`linearisation' of membrane dynamics, when viewed as `moments of continuous
mass in interaction'.  Finally, a note is added which mentions two infinite
classes of minimal hypersurfaces, one of which corresponds to self similarly
expanding, or contracting, open membranes of very intricate
shapes.
\vskip6pt
Let me start by showing how the usual parametric equations,
$$ \partial_\alpha \sqrt{G}G^{\alpha \beta}\partial_\beta x^\mu
= 0 \eqno (1) $$
$$ G_{\alpha \beta}:= {{\partial x^\mu } \over {\partial \varphi^\alpha}}
{{\partial x^\nu} \over {\partial \varphi^\beta}} \eta_{\mu \nu } ,
\hskip1cm \eta_{\mu \nu} = diag (1, -1, ..., -1) $$
$$ \alpha \beta = 0 ... M \hskip1cm  \mu \nu = 0 ... D-1 $$
$$ G = det (G_{\alpha \beta}) \hskip1cm
G^{\alpha \beta}G_{\beta \gamma} = \delta^\alpha _\gamma $$

simplify in the orthonormal light cone gauge

$$\eqalignno{ \varphi^0 &= {1\over 2} (x^0 + x^3) &(2) \cr
\partial_r \zeta &= \dot {\harp x} \partial_r {\harp x} ,
\hskip1cm r = 1...M, &(3) \cr
2\dot {\zeta} &= \dot {\harp x}^2 + g &(4) \cr } $$

(cp. [3]), where $ \zeta : = x^0 - x^3,
{\harp{x}} = (x^1, x^2, x^4, ..., x^{D-1}) $,
. denotes differentiation with respect to $ \varphi^0 =  {{x^0 + x^3} \over 2}
$
(called `time', $t$, in the following argument) and $g$ is the
determinant of the $M \times M$ matrix formed by $g_{rs} : = \partial_r
{\harp{x}}\partial_s{\harp{x}}. $
\vskip6pt
The second order differential operator acting on $x^\mu $ (in (1)) then becomes
$$ D = \partial^2_t - \partial_rgg^{rs}\partial_s \eqno (5), $$
$g^{rs}g_{sr^\prime} = \delta^r_{r^\prime}$, and it is straightforward to show
that
$$ D{\harp{x}} = 0 \eqno (6) $$
implies $D\zeta = 0$, as well as the consistency of $(4)$ with each of (3).
As $Dt$ is automatically zero (cp. (4)), the only remaining condition is the
consistency of (3), i.e.
$$ \partial_r\dot{\harp{x}} \partial_s {\harp{x}}
- \partial_s\dot{\harp{x}} \partial_r {\harp{x}}
= 0 \hskip1cm r,s=1...M \eqno (7). $$
\vskip6pt
So far for minimal surfaces of arbitrary dimension and co-dimension.
\vskip6pt
For $M=2, D=4$, two simple ways of satisfying (7) are:
$$ {\harp{x}} =
\left(\matrix {x(t)\cdot X(\mu ,\varphi )\cr
y(t)\cdot Y(\mu ,\varphi )} \right) \eqno (8), $$
or, assuming rotational symmetry around the $x_3$ - axis,
$$ {\harp{x}} = R(t,\mu)
\left(\matrix {\cos \varphi\cr \sin \varphi } \right)
\hskip1cm \varphi \epsilon [0,2\pi ] \eqno (9) $$
($\mu$ and $\varphi$ denoting the two spatial membrane parameters).  The
dynamical
equations, $D{\harp{x}} = 0$, will be satisfied provided [3]
$$ 2 {\ddot R} = R (R^2)^{\prime\prime} \eqno (10) $$
($ ^\prime = {\partial \over {\partial \mu}}$), respectively
$$ {\ddot x} = -xy^2 , \hskip1cm {\ddot y} = -yx^2 \eqno (11) $$
together with
$$ X = \lbrace \{ X,Y\},Y \rbrace ,
\hskip1cm Y = \lbrace \{ Y,X\}, X\rbrace \eqno (12) $$
($\{ X,Y\}$ denoting $\partial_\mu X \partial_\varphi Y -
\partial_\varphi X \partial_\mu Y$ ) - which basically has only one type of
solution,
$$\eqalignno{
X &= \sqrt {1-\mu^2} \cos \varphi,
\hskip1cm \mu \epsilon [-1, +1] &(13). \cr
Y &= \sqrt {1-\mu^2} \sin \varphi \cr }$$
(11) describes the motion of a unit mass point particle in 2 dimensions,
moving under the influence of the potential $V(x,y) = {1\over 2} x^2y^2$,
a problem which has quite an interesting history (see e.g. [4]-[7]).
Constructing $\zeta$ from (3) and (4) one finds for (8) (with (11)/(13))
$$ \zeta (t,\mu ,\varphi ) = {{1-\mu^2}\over 2} (x\dot x \cos^2\varphi +
y\dot y \sin^2 \varphi ) $$
$$ + {1\over 2} \int^t_0 x^2(\tau )y^2(\tau )d\tau + \zeta_0 \eqno (14) $$
which (upon $ \zeta = x^0 - x^3, t = {{x^0 + x^3}\over 2} $) gives an implicit
equation for $x^3$ as a function of $x^0, \mu$ and $\varphi$, respectively
$x^0, x^1$ and $x^2$ (when using that the $\mu ,\varphi$ dependant term in (14)
is equal to ${1\over 2} (x^2_1 {\dot x\over x} + x^2_2 {\dot y\over y})).$
\vskip6pt
However, even in the simple case
$$ x(t) = y(t) = (2E)^{1\over 4} cn ((2E)^{1\over 4} (t-t_0) ; {1\over {\sqrt
2}})
\eqno (15) $$
it seems difficult to explicitly solve for $x_3$.  The only immediately
tractable case, $y(t) \equiv 0$ and $x(t) = at + b$, giving
$x^2_1 + (x_3 - {b\over a})^2 = (x_0 + {b\over a})^2$ for $a\not= 0$ (a
contracting, or expanding,
circle), and $x^3 = x^0$ (a rigid piece of string, of length $2\mid b\mid$ ,
moving with the velocity of light) for $a = 0$, is of course a `fake' solution,
as $G \equiv 0$.
\vskip6pt
 From (10), on the other hand, one may obtain a simple solution which is both
3 dimensional, and algebraic, by letting
$$ R (t,\mu ) = \pm \sqrt 2 {\sqrt {\mu^2 + e^2}\over t} \eqno (16).$$
 From (3)/(4) one gets
$$ \zeta = - {(\mu^2 + {1\over 3}e^2) \over t^3} \eqno (17), $$
which together with $x^2_1 + x^2_2 = R^2 = 2 {{\mu^2 + e^2}\over t^2}$ gives
$$ (x^0 + x^3)^2 (x^2_0 - x^2_3 + x^2_1 + x^2_2) = {16\over 3}e^2 \eqno (18).
$$
Note that for $\mu \epsilon (-\infty , +\infty)$, (10) has two independant
scaling symmetries, which can be used to look for solutions
$$ R (t,\mu ) = t^a f (t^b \mu ),
\hskip1cm a + b + 1 = 0, \eqno (19), $$
$$ a(a-1) f(s) + sf^\prime (s)(a+1)(2-a) +
(a+1)^2 s^2 f^{\prime\prime} (s) \eqno (20) $$
$$ = (f^{\prime 2} + f f^{\prime\prime} ) f(s) $$
(the case of $a=-1, b=0,$ corresponding to (18)).
\vskip6pt
Let us now look at various other (cp. [1],[2]) reformulations, and
reductions, of (1):
\vskip6pt
Choosing $\varphi^0 = {{x^0 + x^3} \over 2}$ (called $t$, as above),
$\varphi^1 = x^1, \varphi^2 = x^2$ (called $x$ and $y$, respectively) one
obtains a `non-parametric light-cone' or `hydrodynamic'equation for
$x^0 - x^3 =: p(t,x,y)$,
$$ \ddot p + 2 ({\harp{\nabla}}p .
{\harp{\nabla}}\dot p - \dot p {\harp{\nabla}}^2p)
+ {1\over 2} {\harp{\nabla}}p . {\harp{\nabla}}
({\harp{\nabla}}p)^2 - ({\harp{\nabla}}p)^2
{\harp{\nabla}}^2p = 0 \eqno (21), $$
while getting the `usual non-parametric', or `Born-Infeld' - equation,
$$ (1 - z^\alpha z_\alpha) \square z + z^\alpha z^\beta z_{\alpha \beta} = 0
\eqno (22) $$
$$ z_\alpha := {\partial z \over {\partial x^\alpha}},
\hskip.5cm z_{\alpha \beta} = {\partial^2 z \over
{\partial x^\alpha \partial x^\beta}} ,
\hskip.5cm \square z = z_{\alpha \beta}\eta^{\alpha \beta}$$
$$ \eta^{\alpha \beta} = (^1\scriptstyle -1_{\scriptstyle -1}\displaystyle ) $$
for $x^3 = z(x^0, x^1, x^2)$ by choosing $\varphi^0 = x^0, \varphi^1 = x^1,
\varphi^2 = x^2$
(which of course
is appropriate only if the hypersurface can be represented as a graph).  In
the first gauge, $G = 2\dot p + ({\harp{\nabla}}p)^2$ while
$G = 1 - \dot z^2 + ({\harp{\nabla}}z)^2$ in the second
(here, $\dot z: = {\partial z\over \partial x^0}$).
\vskip6pt
Also, one could represent the hypersurface as the set of zeroes (more
generally: a level set) of some scalar function $u$; the dynamical equation
for $u$ that guarantees extremality of the volume $\int \sqrt G$ is
$$ u^\rho u_\rho \square u - u^\mu u^\nu u_{\mu \nu} = 0 \eqno (23) $$
$$ (u_\rho = {\partial u \over {\partial x^\rho}}, \rho = 0...3; u_{\mu v} =
{\partial^2u \over {\partial x^\mu \partial x^v}}, \square u =
u_{\mu \nu} \eta^{\mu \nu}). $$
Assuming the hypersurface to be rotationally symmetric around the $x_3$-axis
(cp. (9)) one obtains
$$ \ddot p + 2 (p^\prime \dot p^\prime - \dot p p^{\prime\prime} ) = {1\over r}
(p^{\prime 3} + 2\dot p p^\prime ) \eqno (24), $$
$$ \ddot z (1 + z^{\prime 2}) - z^{\prime\prime} (1 - \dot z^2) -
2 \dot z z^\prime \dot z^\prime = {1\over r}
(z^{\prime 3} + z^\prime (1 - \dot z^2)) \eqno (25) $$
from (21), respectively (22).  Note that both forms look considerably more
complicated than (10), which corresponds to the Lagrangian density
$L = {1\over 2} (\dot R^2 - R^2 R^{\prime 2})$ or - with
$\phi = {1\over 2} R^2$ - a Hamiltonian density
$H = \phi \pi^2 + {1\over 2} \phi^{\prime 2}$.
The relation with strings in a curved 3 dimensional background is easily seen
by noting that the action for (25),
$S = \int r dr dt \sqrt {1 - \dot z^2 + z^{\prime 2}}$, is a non-parametric
version of $S = \int d\varphi^0 d\varphi^1 \sqrt {-det (\partial_a x^\alpha
\partial_b x^\beta g_{\alpha \beta}})$, with
$g_{\alpha \beta} = r\eta^{\alpha \beta}, ds^2 = r(dt^2 - dr^2 - dz^2)$
(implying a curvature singularity according to $R = - {3\over {2r^3}}$).
\vskip6pt
Further reducing (24) by letting
$$ p (t,r = \sqrt {x^2_1 + x^2_2}) = t^a P (t^c r = Z) \eqno (26) $$
$$ a + 2c = 1 = 0 $$
one gets
$$ a(a - 1)P - {3\over 4} (a^2 - 1) Z P^\prime +
{1\over 4} (a + 1)^2 Z^2 P^{\prime\prime} - {P^{\prime 3} \over Z} +
2a  (P^{\prime 2} - {PP^\prime \over Z} - PP^{\prime\prime}) = 0 \eqno (27). $$
At least the case of $a = 0$ can be solved by elementary methods, yielding
an elliptic integral, respectively
$$ x^3 - x^0 = \pm {1\over 2} \int^{2({x^2_1 + x^2_2 \over {x^0 + x^3}})}_0
{du \over {\sqrt {1 - ({u\over u_0})^4}}} + const. \eqno (28) $$
A rather large class of solutions can be obtained from the Ansatz
$$ z (x^0, x^1, x^2) = x^0 - p(x^1, x^2) \eqno (29), $$
respectively $\dot p = 0$ in (21), yielding the equation
$$ p^2_x p_{yy} + p^2_y p_{xx} - 2p_x p_y p_{xy} = 0 \eqno (30) $$
for the `shape of the surface that moves with the velocity of light in the
$x^3$-direction'.  While the integrability of (30) must have been known for
quite a long time, the above connection with extremal hypersurfaces in
Minkowski space was noted only recently, in collaboration with M. Bordemann.
Instead of resorting to the general method of linearisation by hodograph -
transform (applicable to any 2-dimensional field equation that comes
from a lagrangean which depends only on the first derivatives of the field)
solutions of (30) may actually be obtained in rather more explicit form,
which is quite useful for a qualitative discussion in the membrane context.
It is e.g. easy to show that
$$ p(x,y) : = \tilde p (x + v(x,y) y) \eqno (31), $$
with
$$ v(x,y) = \tilde v (x + v(x,y) y) \eqno (32), $$
solves (30), for any smooth $\tilde p$ and $\tilde v$.  As (32) implies
$$ v_x = {\tilde v^\prime \over {1 -y\tilde v^\prime}},
v_y = {v\tilde v^\prime \over {1-y\tilde v^\prime}} \eqno (33),$$
$v$ and $p$ will generically have `cusps' in the ($xy$) plane, as well as
(when moved according to (29)) somewhere vanishing $G$ (both, however, only
on measure zero sets), according to
$$ G = (\tilde p^\prime)^2 {(1 + v^2) \over {(1 - y\tilde v^\prime)^2}} \eqno
(34). $$
Note that for $\tilde v$ = const. any strictly montonic $\tilde p$ determines
a hypersurface which is everywhere regular.
In a different context, Fairlie et al [8] (partly referring to work of Bateman
and Garabedian) have discussed various aspects of (30), including the existence
of solutions defined in terms of two functions, F and H, via
$xF(p) + yH(p) \equiv 1$.  When moved according to (29), the corresponding
hypersurfaces will have $G = {F^2 + H^2 \over ({xF^\prime + {yH^\prime )^2}}}$.
Similarly, $p = p (\cot^{-1} ({x\over y}))$ will give
$\int \sqrt G \sim \int^{2\pi}_0 d\theta \mid p^\prime (\theta ) \mid.$
\vskip6pt
Finally note that (22) may be viewed as a consistency condition of a free
wave equation in 4 dimensions,
$$h^{\mu \nu}\partial^2_{\mu \nu} z = 0, \hskip1cm
h^{\mu \nu} = diag (1, -1, -1, +1) \eqno (35)$$
and a non-linear constraint,
$$ h^{\mu \nu}\partial_\mu z \partial_\nu z = 1 \eqno (36).$$
In light cone coordinates, $r = {x^0 + x^2 \over 2}, s = {x^0 - x^2\over 2},
u = {x^1 + x^3\over 2}, v = {x^1 - x^3\over 2}$ the two equations read
$$f_{rs} = f_{uv}, \hskip1cm f_rf_s - f_uf_v = 1 \eqno (37).$$
The simplest solutions obtained this way are, $\pm f = r + s + g (u$ or $v)$,
respectively $u - v + g (r$ or $s)$.
\vskip6pt
Let me conclude by mentioning a kind of `linearization' that occurs when
looking at eq. (23) in the following way (I thank Martin Bordemann for many
discussions on this direction):
A field dependant change of variables from $x^\mu$ to $p_\mu : =
{\partial u \over {\partial x^\mu}}$ transforms
$S = \int d^4 x  \sqrt {\partial_\mu u \partial^\mu u}$ into
$S = \int d^4p \sqrt {p^\mu p_\mu} det M(p)$, where $M$ is the matrix of
second derivatives of a scalar field $f(p_0, ..., p_3)$.  This suggests taking
$\sqrt {p^\mu p_\mu}$ as one of the independent variables, ie. foliating the
relevant part of flat Minkowski $p$-space according to
$p^\mu = r e^\mu ({\harp q}), e^\mu e_\mu = 1 \
({\harp q} = (q^1, q^2, q^3)$ parametrizing the hyperboloid
$H^3$).  Due to the specific nature of (23), terms explicitly depending on the
independent coordinates cancel in the equation of motion for
$f = f (r, {\harp q})$, which simply becomes
$$\epsilon^{i\mu_1\mu_2\mu_3} \epsilon_i\ ^{\nu_1\nu_2\nu_3} f;_{\mu_1\nu_1}
f;_{\mu_2\nu_2} f;_{\mu_3\nu_3} = 0 \eqno (38),$$
; indicating covariant differentiation with respect to the induced metric,
ie.
$f;_{00} = f_{rr}, f;_{0i} = f_{ri} - {1\over r}f_i, f;_{ij} =
f_{ij} - rg_{ij} f_r$
($g_{ij}$ being the metric on $H^3$, eg.
${4\over {(1 - {\harp q}^2)^2}} \delta_{ij}$ in the `Poincar\'e-ball' model,
$0 \leq \mid {\harp q} \mid \leq 1$).
\vskip6pt
For $f(r,{\harp q}) = \sum^\infty_0 r^n Y^{(n)}{(\harp q)}, (38)$
becomes a system of non-linear differential equations for the set
$\{ Y^{(n)}\} $ of scalar functions on $H^3$.  However, the equation for
$Y^{(0)}$ is just the level-set version for minimal surfaces in $H^3$
(cp. (23), or [2]; and [9, 10, 12] for the integrable nature of this problem,
and
examples), while each $Y^{(n)}, n > 0$, may recursively be determined in
terms of the $Y^{(m < n)}$ as a solution of a linear(!) equation.
\vskip6pt
{\bf Acknowledgement:}
\vskip6pt
I would like to thank the organisers and participants of the `Geometry and
Gravity' workshop for the stimulating atmosphere and discussions; in
particular, I am grateful to S.T. Yau for his interest.
\vfill\eject
{\bf Note added:}
\vskip6pt
Let [11]
$$e : \Sigma_2 \rightarrow H^3 : = \{ e^\mu \epsilon R^{1,3} \mid
e^\mu e_\mu = + 1 \} \eqno (39),$$
resp.
$$\tilde e: \tilde \Sigma_2 \rightarrow S^{2,1} = \{ \tilde e^\mu
\epsilon R^{1,3} \mid
\tilde e^\mu \tilde e_\mu = -1 \} \eqno (40),$$
be a spacelike (rsp. timelike) 2 dimensional surface with zero mean curvature
in $H^3$ (resp. $S^{2,1}$).  The cone $C(\Sigma )$ (resp. $\tilde C
(\tilde \Sigma )$),
defined by
$$ X : (0, \infty ) \times \Sigma_2 \rightarrow R^{1,3} \eqno (41),$$
$$ \hskip2truecm \rho, (\varphi^1 \varphi^2 ) \rightarrow \rho \cdot
e(\varphi^1 \varphi^2 )$$
resp.
$$ \eqalignno{\tilde X : \tilde \Sigma_2 &\times (0, \infty ) \rightarrow
R^{1,3} \cr
&(\varphi^0 \varphi^1), \tilde \rho \rightarrow \tilde \rho \cdot \tilde e
(\varphi^0 \varphi^1) &(42) \cr }$$
is then a 3 dimensional timelike hypersurface with zero mean curvature in
$R^{1,3}$ [11].
\vskip6pt
This is easy to check, as eg. (1) for $X$ (resp. $\tilde X$), with
$\varphi^0 : = \rho $
(implying $G_{00} = + 1, G_{0r} = 0, G_{rs} = \rho^2 h_{rs}$, with
$h_{rs} : = \partial_r e^\mu \partial_s e_\mu$ negative definite; $r,s = 1,2$)
- respectively $\varphi^2 : = \tilde \rho$ (implying $G_{22} = -1, G_{a2} = 0,
G_{ab} = \tilde \rho^2 \tilde h_{ab},  {\sqrt G} = \tilde \rho^2 {\sqrt
{-\tilde h}}$,
where $\tilde h$ is the
determinant of $\tilde h_{ab} : = \partial_a \tilde e^\mu \partial_b
\tilde e_\mu ;
a,b = 0,1$) reads
$$ (2 + {1\over {\sqrt h}} \partial_r {\sqrt h} h^{rs} \partial_s ) e^\mu = 0
\eqno (43),$$
resp.
$$ (-2 + {1\over {\sqrt {-\tilde h}}} \partial^a {\sqrt {-\tilde h}}
\tilde h^{ab} \partial_b) \tilde e^\mu = 0 \eqno (44). $$
Correspondingly, one may eg. choose $\varphi^0 : = x^0$, and consider the
Ansatz
$$ {\harp x} (t) = t\cdot {\harp v} (\varphi^1, \varphi^2) \eqno (45), $$
$\mid {\harp v} \mid \leq 1$.  The resulting equation for ${\harp v}$ is
the minimal surface equation in the `Klein-ball' model of $H^3$.
As much is known about minimal surfaces in hyperbolic 3-space [9, 10, 12]*
(basically, the problem reduces to solving variants of the integrable
Sinh-Gordon equation), infinitely many possible membrane shapes, expanding
according to (45), may thus be determined.
Finally, note again the relation with strings in curved 3 dimensional
backgrounds (via (40), or (10)/(23), eg.), especially with regard to a quantum
theory of relativistic surfaces.
\vfill\eject
{\bf References}
\vskip6pt
\item{[1]} M Bordemann, J Hoppe; Phys Lett B 317 (1993) 315.
\vskip6pt
\item{[2]} M Bordemann, J Hoppe; `The Dynamics of Relativistic Membranes II
Nonlinear Waves and Covariantly Reduced Membrane Equations' FR THEP-93-19,
KA-THEP-5-93, August 1993.
\vskip6pt
\item{[3]} J Hoppe; `Quantum Theory of a Massless Relativistic Surface and
...', MIT PhD Thesis (1982); Elem Part Res J (Kyoto) 80 (1989) 145.
\vskip6pt
\item{[4]} J Hoppe; `Two Problems in Quantum Mechanics', MIT MSc Thesis (1980).
\vskip6pt
\item{[5]} B Simon; Annals of Physics 146 (1983) 209.
\item{} C Feffermann; Bul Am Math Soc Vol 9 \#2 (1983) 129.
\vskip6pt
\item{[6]} SL Ziglin; Funct Anal Appl 17 (1981) 6.
\item{} ES Nikolaevskii, LN Schur; JETP Lett 36 (1982) 218.
\vskip6pt
\item{[7]} P Dahlquist, G Russberg; Phys Rev Letters 65 (1990) 2837.
\vskip6pt
\item{[8]} DB Fairlie, J Govaerts, A Morosov; Nucl Phys B 373 (1992) 214.
\vskip6pt
\item{[9]} M Babich, A Bobenko; Duke Math J Vol 72 (1993) 1.
\vskip6pt
\item{[10]} K Polthier; `Geometric A Priori Estimates for Hyperbolic Minimal
Surfaces', SFB 256 preprint \# 300, 1993, Bonn University.
\vskip6pt
\item{[11]} B Palmer; private communication.
\vskip6pt
\item{[*]} Many thanks to A Bobenko, B Palmer, U Pinkall and K Polthier for
valuable discussions.
\vskip6pt
\item{[12]} H Wente; `Constant Mean Curvature Immersions of Enneper Type',
Mem Am Math Soc, 1992.
\bye